\documentclass[10pt]{article}
\usepackage{spconf,amsmath,graphicx}
\usepackage{verbatim}
\usepackage{amsfonts}
 \usepackage{dsfont}
\newcommand*{\bigchi}{\mbox{\large$\chi$}}
\usepackage{enumitem}

\title{Multiscale Community Mining in Networks\\ Using Spectral Graph Wavelets}
\name{Nicolas Tremblay\thanks{This work was partially supported by the GDR ISIS and the CNRS/JSPS Grants ``Benchmarking and statistical analysis tools for modern Internet and sensor network traffics,'' 2010 and 2012.}, Pierre Borgnat}
\address{CNRS, Laboratoire de Physique (UMR 5672), \'Ecole Normale Sup\'erieure de Lyon, France}

\begin{document}
%
\maketitle
\begin{abstract}
For data represented by networks, the community structure of the underlying graph is of great interest. 
A classical clustering problem is to uncover the overall ``best'' partition of nodes in communities. Here, a more elaborate description is proposed in which community structures are identified at different scales. 
To this end, we take advantage of the local and scale-dependent information encoded in graph wavelets. 
After new developments for the practical use of graph wavelets, studying proper scale boundaries and parameters and introducing scaling functions, we propose
a method to mine for communities in complex networks in a scale-dependent manner. It relies on classifying nodes according to their wavelets or scaling functions, 
using a scale-dependent modularity function. An example on a graph benchmark having 
hierarchical communities shows that we estimate successfully its multiscale structure.
\end{abstract}
\section{Introduction}
\label{sec:intro}

In a large number of applications, data are naturally represented as networks (or weighted graphs): social networks, sensor networks, Internet networks, neuronal networks, transportation networks, biological networks,... Analyzing such networks has been emerging in the last decade as a central 
issue in the study of complex networks~\cite{Chung2006, Barrat2008}.  
A striking property of many networks, and a common way of simplifying the network's analysis, is their modular structure, i.e., there exists groups of nodes that are more connected with themselves than with the rest of the network; these groups are called communities~\cite{Fortunato2010}. 
As nodes in a same community tend to share common properties, community mining provides both a sketch of the structure of a network, and some insight on nodes' properties. 
One issue in community mining is defining the scale at which one wants to analyze the network. 
Many algorithms  (see the review~\cite{Fortunato2010}) are based on the optimisation of appropriate evaluation functions 
such as the popular modularity~\cite{newman2006modularity} and generally discard this question or propose only \emph{ad-hoc} discussions. 
Modularity for instance is known to favour an intrinsic scale of description~\cite{fortunato2007resolution,kumpula2007limited}.
 
The present work develops a scale-dependent procedure which 
identifies community structures, i.e., that classifies nodes according to their community, at different scales.
After one decides on a scale of interest
(or a collection of scales), the objective is to mine for the communities at this(ese) scale(s).
Using multiple scales will provide a fully multiscale community description of the network.
Some authors have proposed multiscale community mining either based on 
random walk processes~\cite{schaub2012markov,lambiotte2010multi}, or on definitions of parametric modularities \cite{reichardt2006statistical,arenas2008analysis}. 
Our proposition for community mining is to rely on the recent construction of graph wavelets based on spectral graph theory~\cite{hammond2011wavelets}.

By nature, the wavelet associated to a node $a$ and a scale $s$ is local. It is centered around this node and spreads on its neighbourhood
so that the larger is $s$, the larger is the spanned neighbourhood. In some sense, wavelets give an ``egocentered'' 
view of how a node ``sees'' the network at that scale. 
Taking advantage of this local information encoded in wavelets, we develop an approach that clusters together nodes whose local environments are similar, i.e., whose associated wavelets are correlated. 
Then, to uncover the community structure at a given scale, we circumvent the intrinsic scale issue of the classical modularity by defining 
a scale-dependent modularity (using wavelets) which leads to this scale's community structure when it is maximized.
While graph wavelets have been applied now to classical problems in signal processing (e.g., source estimation for EEG~\cite{hammond2012incorporating}) 
or image processing~\cite{Narang2012}, it has never been used for the detection of community in networks.



Section \ref{sec:wavelets} recalls useful background material: an example of network with multiscale communities, and  elements of spectral graph theory and wavelets.
Some contributions to the use of spectral graph wavelets are presented in section \ref{sec:3}: in \ref{sec:3p1}, we discuss that once a network is given, a proper choice of scale boundaries ends up with parameters for the band-pass filter defining the wavelets that are different from~\cite{hammond2011wavelets}. 
In \ref{sec:scf}, scaling functions are defined for graph wavelets, as they will prove more robust than wavelets for community mining
(up to our knowledge, this is the first introduction of scaling functions for graph wavelets).
In section \ref{sec:detection}, the algorithm for multiscale community mining is described and it is applied on the benchmark in section \ref{sec:examples}.
We conclude in section \ref{sec:concl}.

\section{Background Material}
\label{sec:wavelets}

The present work relies on one side on the notion of multiscale communities in graphs, as discussed for instance in \cite{lambiotte2010multi,sales2007extracting}, and on spectral graph wavelets as introduced in \cite{hammond2011wavelets} on another. The reader is supposed to have some familiarity with both works.
Still, this section recalls useful elements.

\subsection{Model of network with multiscale communities}
\label{sec:2model}
Following~\cite{lambiotte2010multi}, the model of graph defined in~\cite{sales2007extracting}, for  which the multiscale community structure is known, is adopted as a benchmark to test multiscale community mining.
The global density of links is controlled by a first parameter $\bar{k}$ (which is the mean degree of the nodes). Then,
the intra-community and inter-community relative density of links is fixed by a second parameter $\rho$. 
As an example, we choose $\bar{k}=16$ and $\rho=1$. We consider 
graphs of 640 nodes, divided in three  hierarchical levels: there are 64 small communities of 10 nodes each (the finest scale) embedded in 16 communities of 40 nodes each (the intermediate scale), 
themselves embedded in 4 communities of 160 nodes each (the coarsest scale). 
A realization of the network is visualized in Fig.~\ref{fig:graph}.

\subsection{Spectral Graph Theory and Fourier Modes}
Let $\mathcal{G}=(V,E,A)$ be a undirected weighted graph with $V$ the set of nodes, $E$ the set of edges, and $A$ the weighted adjacency matrix such that $A_{ij}=A_{ji}\geq0$ is 
the weight of the edge between nodes $i$ and $j$.  Note $N$ the total number of nodes. 

Let us define the graph's Laplacian matrix $L=D-A$ where $D$ is a diagonal matrix 
with $D_{ii}= d_i = \sum_{j\neq i} A_{ij}$ the strength of node $i$.
$L$ is real symmetric, therefore diagonalisable: its spectrum is composed of its sorted eigenvalues $\left(\lambda_l\right)_{l=0\dots N-1}$, so that
$\lambda_0\leq\lambda_1\leq\lambda_2\leq\dots\leq\lambda_{N-1}$; and of the matrix $\mathbf{\bigchi}$ of its normalized eigenvectors:
$\mathbf{\bigchi}=\left(\chi_0|\chi_1|\dots|\chi_{N-1}\right)$. 
Considering only connected graphs, the multiplicity of eigenvalue $\lambda_0=0$ is 1. 

By analogy to the continuous Laplacian operator whose eigenfunctions are the continuous Fourier modes and eigenvalues their 
squared frequencies, $\mathbf{\bigchi}$ is considered as the matrix of the graph's Fourier modes, and $\left(\sqrt{\lambda_l}\right)_{l=0\dots N-1}$ its set of associated 
``frequencies''. More details are found also in \cite{hammond2011wavelets}.

\subsection{The Graph Wavelets}
\label{subsec:wavelets}
Graph wavelets were defined in \cite{hammond2011wavelets} using the graph Fourier modes. 
Let us note $\psi_{s,a}$ the wavelet centered around node $a$.
Its construction is based on band-pass filters defined in the graph Fourier domain, 
generated by stretching a band-pass filter kernel $g(\cdot)$ by a scale parameter $s>0$. 
The stretched filter has matrix representation $\hat{G_s}=diag(g(s\lambda_0), \dots, g(s\lambda_{N-1}))$ 
that is diagonal on the Fourier modes (the $N$ eigenvectors of $L$).
Hence, the wavelet basis at scale $s$ reads as: 
$$\Psi_s=\left(\psi_{s,0}|\psi_{s,1}|\dots|\psi_{s,N-1}\right)
= \bigchi\hat{G_s}\bigchi^\top.$$
Here, our main use is of the localized wavelets $\psi_{s,a}$ themselves.
Note that the wavelet transform at scale $s$ of a signal $f$ would be obtained by decomposing $f$ on $\Psi_s$.

The intuition behind this definition of wavelets on graphs is that, at small scales (small scale parameter $s$), the filter $g(s \, \cdot)$ is stretched out, thus letting through 
high frequency modes essential to good localization: corresponding wavelets extends only to their close neighbourhood in the graph. At large scales 
(large $s$) the filter function is compressed around low frequency modes and this creates wavelets encoding a coarser description of the local environment.

\label{subsec:sc bnd}
We use the band-pass filter kernel $g$ proposed in \cite{hammond2011wavelets}:
\begin{equation}
\label{eq:BPfilter}
  g (x; \alpha , \beta, x_1 , x_2 ) =
  \left\{
      \begin{aligned}
        &x_1^{-\alpha}\,x^\alpha &\mbox{for}\quad x<x_1\\
        &p(x) &\mbox{for}\quad x_1\leq x\leq x_2\\
        &x_2^{\beta}\,x^{-\beta} &\mbox{for}\quad x>x_2.\\
      \end{aligned}
    \right.
\end{equation}
$p(x)$ is taken as  the unique cubic polynomial interpolation that respects the continuity of $g$ and its derivative $g'$. 

We propose in the following a new way to set the parameters different from ~\cite{hammond2011wavelets}, by studying in \ref{sec:3p1} the scale boundaries that are relevant for community mining.

\begin{figure}[t]
\centerline{\includegraphics[height=4.9cm]{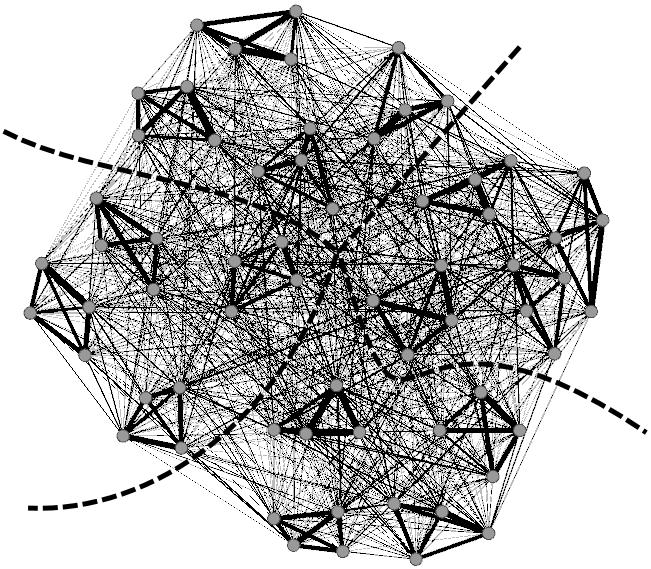}}
\caption{Sketch of the graph discussed in \ref{sec:2model}: 
each node displayed is actually a community of 10 nodes. 
The thickness of each link is proportional to the total number of links between the
two corresponding communities.  
}
\label{fig:graph}
\end{figure}

\section{Graph wavelet parameters and graph Scaling function for community mining}
\label{sec:3}


We first propose a new way to specify the parameters of $g$ defining the graph wavelets, by studying the range for the scales $s$ relevant for community mining. Then, we introduce graph scaling functions associated to graph wavelets.

\subsection{Band-pass filter parameters and scale range}
\label{sec:3p1}
For the band-pass filter kernel $g$ of eq.~(\ref{eq:BPfilter}), parameters $\alpha, \beta$, $x_1$ and $x_2$ have to be carefully chosen 
so as to generate appropriate wavelets, as well as the scales $s$ to be used. 
Instead of following \cite{hammond2011wavelets}, 
the parameters are based here on an argument of spectral clustering of graphs~\cite{Fortunato2010}: 
the eigenvector $\chi_1$ (associated to the smallest non-zero eigenvalue $\lambda_1$) is the first in importance for community mining 
because it contains information of the coarsest description of the graph.

A first consequence is that the maximum scale parameter $s_{max}$ is set so that the filter function $g(s_{max}\,x)$ starts decaying as 
a power law only after $x=\lambda_1$, hence $s_{max}=x_2/\lambda_1$.
We require also that the filter at the maximum scale is highly selective around $\lambda_1$; for that, all other eigenmodes (especially $\lambda_2$) 
have to be attenuated. 
Choosing an attenuation by a factor 10, this leads us to:
$g(s_{max}\,\lambda_1)=10\,g(s_{max}\,\lambda_2)$, hence $\beta={1}/{\log_{10}\left(\frac{\lambda_2}{\lambda_1}\right)}$.

Second, we need to keep a part of $\chi_1$ in the wavelets of every scale, so that all wavelets  
are sensitive to large scale community structure. 
We propose as minimum scale $s_{min}$ the one for which $g(s_{min}\,\lambda_1)$ becomes 
smaller than $1$. Using eq.~(\ref{eq:BPfilter}), this gives  $s_{min}={x_1}/{\lambda_1}$. 
Imposing also that $g(s_{min}\, \cdot)$ spans the whole range of eigenvalues (so that the wavelet basis is not blind to any eigenvalue), 
we should have $s_{min}\,\lambda_{N-1}=x_2$. 

\begin{figure}[tb]
\begin{minipage}[b]{.48\linewidth}
  \centering
  \centerline{(a) \includegraphics[width=4.0cm]{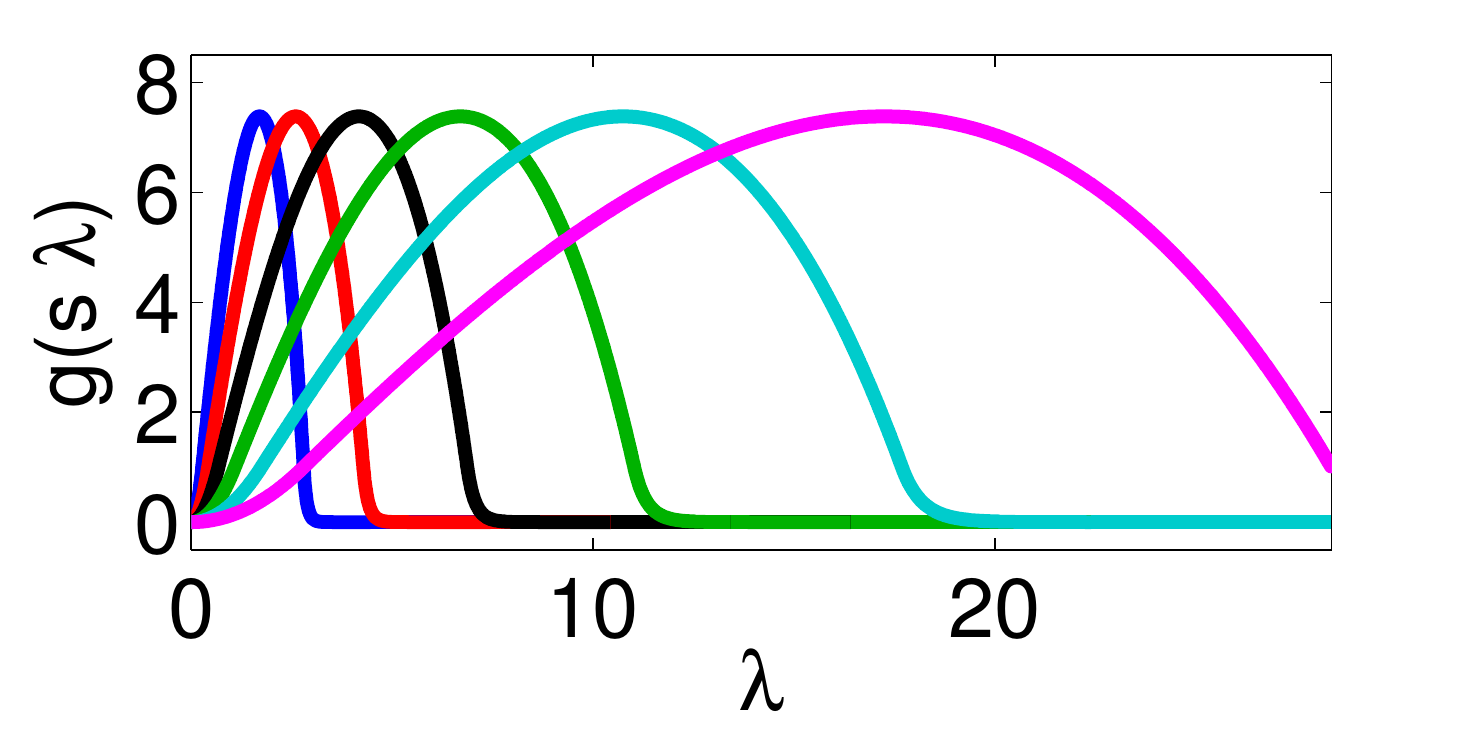}}
\end{minipage}
\hfill
\begin{minipage}[b]{0.48\linewidth}
  \centering
  \centerline{(b) \includegraphics[width=4.0cm]{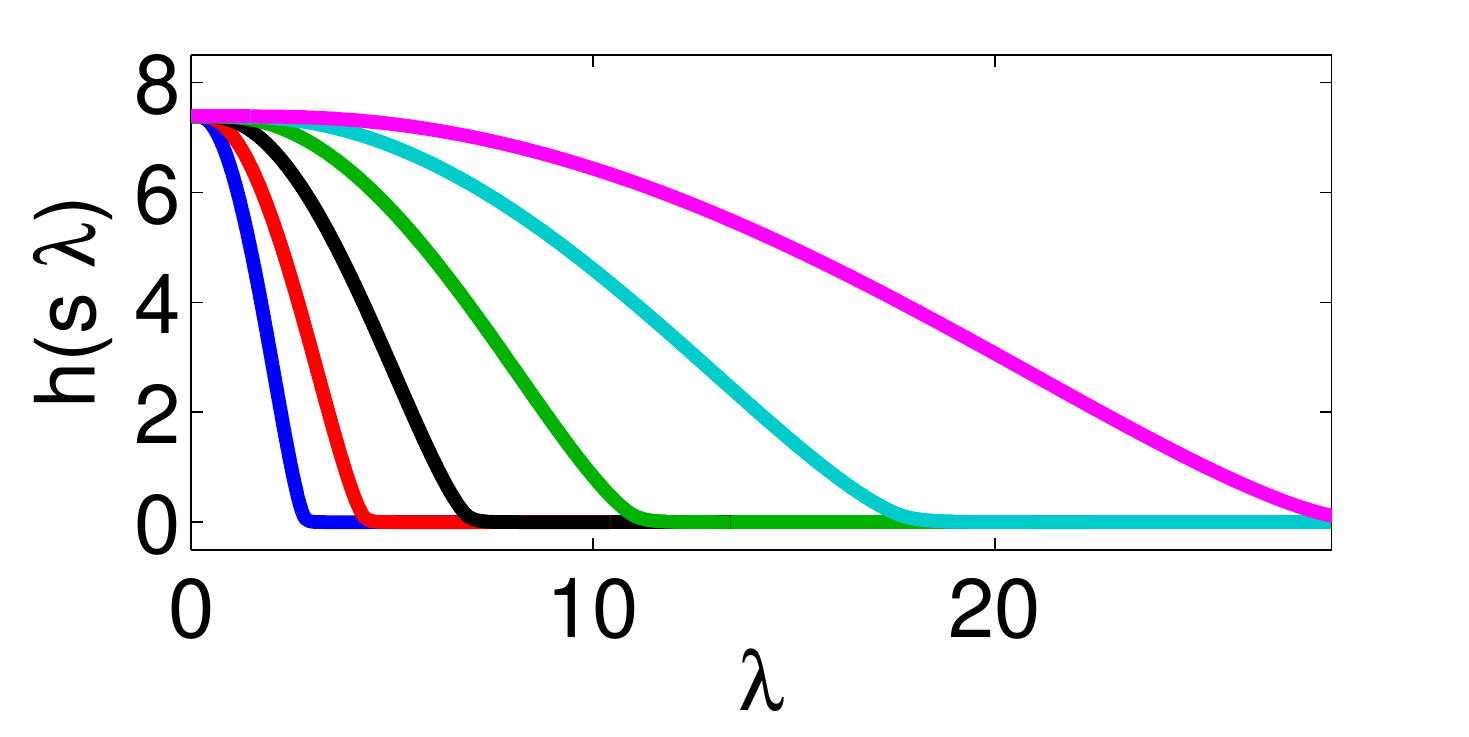}}
\end{minipage}
%

\caption{(a) Band-pass filter functions $g$ and (b) their associated low-pass filter functions $h$ for six different scales within the scale boundaries:  
$s= $ 3.6\,($s_{max}$), 2.4, 1.5, 0.9, 0.6 and 0.4\,($s_{min}$); and the parameters from~\ref{sec:3p1}: $x_1=1$, $x_2=10$, $\alpha=2$ and $\beta=35$.
Filters computed for the network in Fig.\ref{fig:graph}.
}
\label{fig:wavelets}
\end{figure}

This argumentation gives us a value for $\beta$ and three equations linking $x_1$, $x_2$, $s_{min}$ and $s_{max}$.
For $x_1$ and $\alpha$ describing the cut-off at low frequency, there is no argument from spectral clustering to set them. 
Following~\cite{hammond2011wavelets}, we set $x_1=1$ and $\alpha=2$ . This in turn sets $x_2={\lambda_{N-1}}/{\lambda_1}$ and, thereby, 
$s_{max}=\lambda_{N-1}/\lambda_1^2$ and $s_{min}=\frac{1}{\lambda_1}$.
%
%
%
Fig.~\ref{fig:wavelets}.(a) shows examples of band-pass filters $g(s\cdot)$ when using the proposed range of scale and parameters, computed for the graph of Fig.~\ref{fig:graph}.

\subsection{Graph scaling functions}
\label{sec:scf}

Let us introduce scale dependent scaling functions on graphs by analogy to the case of the continuous wavelet transform \cite{CarmonaBook}.
For that, we define the scale-dependent low-pass filters, noted $\hat{H_s}$, derived from a unique low-pass filter kernel $h$ stretched by the scale parameter $s$: 
$$\hat{H_s}=diag(h(s\lambda_0),h(s\lambda_1),\dots,h(s\lambda_{N-1})).$$ 
For consistency with classical wavelet theory \cite{CarmonaBook}, we impose
$$h(x)=\left(\int^\infty_x  \frac{|g(x')|^2}{x'}dx'\right)^{1/2}.$$
Then, the columns $\phi_{s,a}$ of $\Phi_s=\bigchi\hat{H_s}\bigchi^\top$ may be understood as scaling functions at scale $s$.
Fig.~\ref{fig:wavelets}.(b) shows the low-pass filters corresponding to these scaling functions.
These graph scaling functions will be useful in the next sections, as they will lead to better results.

\section{Multiscale Community Mining}
\label{sec:detection}

\subsection{Elements of the method}


\noindent \textbf{Clustering of nodes according to their wavelets.}
For community mining, we aim at grouping together nodes whose environments are similar. 
As the local environment at a given scale $s$ is encoded in the graph wavelets at each node $a$ for this scale, we use $\psi_{s,a}$ as a feature vector 
for the local view from node $a$. A comparison of the views from two nodes $a$ and $b$ is obtained by taking the correlation distance 
(equal to 1 minus the correlation coefficient) between wavelets:
$$
d^g_s(a,b)  = 1 - (\tilde{\psi_{s,a}})^T \tilde{\psi_{s,b}}
$$
where $\tilde{\psi_{s,a}}$ is the wavelet 
after normalization in energy.
%

In order to group nodes together, a hierarchical complete-linkage clustering algorithm~\cite{jain1999data} 
is used, with an additional connectivity constraint~\cite{donetti2004detecting}: a node cannot be clustered in a group of nodes to which it has no path in the original network. 
This algorithm outputs a dendrogram.
As we do not know beforehand how many clusters there are in the network, we have to evaluate each possible 
subdivision of the dendrogram and estimate which one is relevant.

\noindent \textbf{Estimation of the number of communities with a scale-dependent modularity.}
As an evaluation function of the relevance of a given clustering, we introduce a scale-dependent definition of the modularity. 
Let $S$ be a matrix of size $N \times J$ coding for a community clustering:
$S=\left(\mathds{1}_{C_1}|\mathds{1}_{C_2}|\dots|\mathds{1}_{C_J}\right)$ where $\mathds{1}_{C_j}$ is the binary indicator 
function of community $C_j$ (i.e., ${\mathds{1}_{C_j}}_i=1$ if node $i$ is in $C_j$, else 0). 
The classical modularity matrix $B$ is defined as \cite{newman2006modularity}:
$$B=\frac{1}{2m}\left(A-\frac{d d^\top}{2m}\right)$$ where $m$ is the sum of all weights in the graph and $d$ the vector of strengths of the nodes.
The usual modularity is then computed as: $Q={\rm tr}(S^\top BS)$.
Here, we introduce a filtered version of $B$ at each scale $s$:
$$B^g_s=\Phi_s^\top B=\bigchi\hat{G_s}\bigchi^\top B.$$
Then, a scale-dependent modularity is  $Q^g_s={\rm tr}(S^\top B^g_sS)$.

Choosing the subdivision of the dendrogram 
(from the clustering of nodes) that 
maximizes the scale-dependent modularity leads to a community
structure at scale $s$. 
We thereby circumvent the existing problem of classical modularity which imposes its own uncontrolled scale~\cite{fortunato2007resolution,kumpula2007limited}.

In this method, we may replace wavelet functions (with filter $g$) by scaling functions (with filter $h$). In that case, the average of the  scaling functions is removed to compute the correlation distance $d^h_s$. Results with both choices will be shown and we discuss afterwards which one is to be preferred.

\subsection{Proposed method for multiscale community mining}
\label{ssec4:protocol}
The proposed method is summarized as follows.
\begin{enumerate}[noitemsep,topsep=2pt,parsep=0pt,partopsep=0pt]
 \item Choose one scale $s$ in the range of scales $[s_{min}, s_{max}]$ relevant to study the community structure (see \ref{sec:3p1}).
 \item Compute the graph wavelets (resp. the scaling functions) at $s$ for each node
 (as described in \ref{subsec:wavelets}, resp. \ref{sec:scf}).
 \item Cluster the nodes according to the correlation distance $d^g_s$ between wavelets (resp. $d^h_s$ between scaling functions), using a hierarchical complete-linkage clustering. The output is a dendrogram.
 \item Evaluate each subdivision of the dendrogram with the filtered modularity, and
 keep the one yielding the maximum filtered modularity $Q^g_s$ (resp. $Q^h_s$). 
\end{enumerate}
This outputs the community structure at scale $s$. To obtain 
a full multiscale description, repeat the steps with a sampling of the
scales between the scale boundaries.
 
\begin{figure}[t]
\begin{minipage}[b]{.48\linewidth}
  \centering
  \centerline{\includegraphics[width=4.0cm]{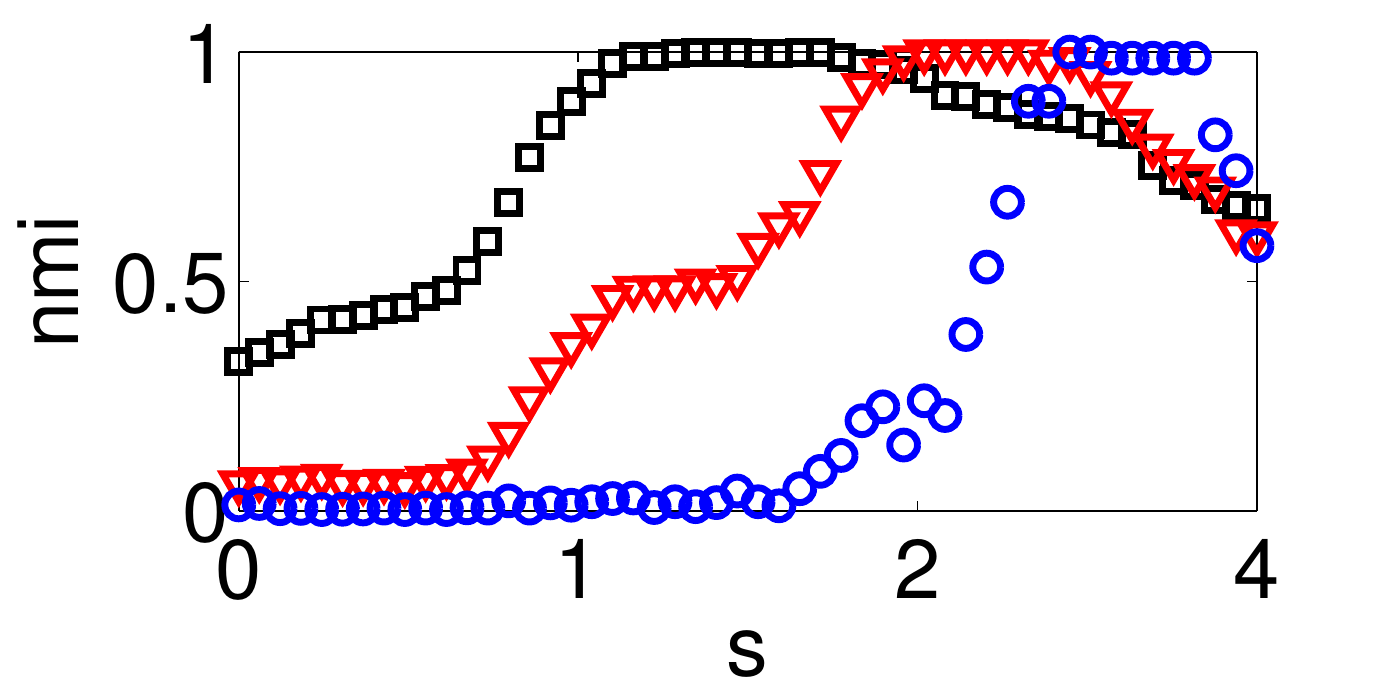}}
\end{minipage}
\hfill
\begin{minipage}[b]{0.48\linewidth}
  \centering
  \centerline{\includegraphics[width=4.0cm]{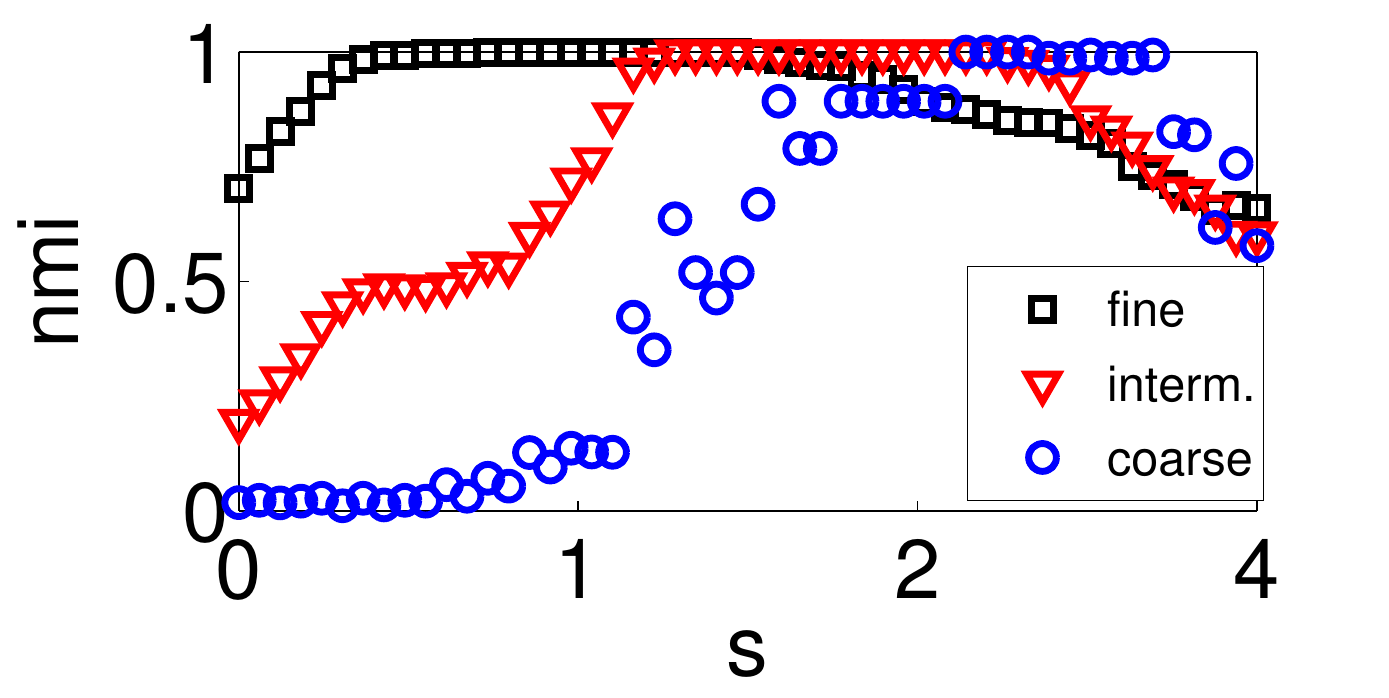}}
\end{minipage}
\caption{
Normalized mutual information ({nmi}) between the theoretical partition and the empirical one obtained when cutting the dendrogram with theoretical number of clusters. Left: using wavelets. Right: using scaling functions.
}
\label{fig:results}
\end{figure}

\section{Test of the method on a Hierarchical benchmark network}
\label{sec:examples}

We test the algorithm on the model of network described in \ref{sec:2model}.
%
%
%
Following the method of \ref{ssec4:protocol}, we obtain in step 3 one dendrogram per scale. 
Let us first assess the validity of the dendrograms.
For each of the three levels of description, we compute the normalized mutual information ({nmi})~\cite{danon2005comparing} between the known 
theoretical partition and the empirical one obtained when cutting the dendrogram with \emph{a priori} 
knowledge of the total number of clusters. We plot it with respect to the scale parameter in Fig.~\ref{fig:results}.
The black squares stands for the fine level of description, the red triangles for the intermediate one, and the blue circles for the coarsest one.
A {nmi} value of 1 means that both partition are the same. 
Results are obtained using wavelets (left), or scaling functions (right).
We see that there always exists an interval of scales for which one may recover 
the theoretical partition from the dendrogram. When using scaling functions, the intervals at value 1 are
larger, especially for small scale. This validates the first three steps of our method: the dendrograms do contain the relevant scale-dependent information. 

\begin{figure}[tb]
\begin{minipage}[b]{.48\linewidth}
  \centering
  \centerline{\includegraphics[width=4.0cm]{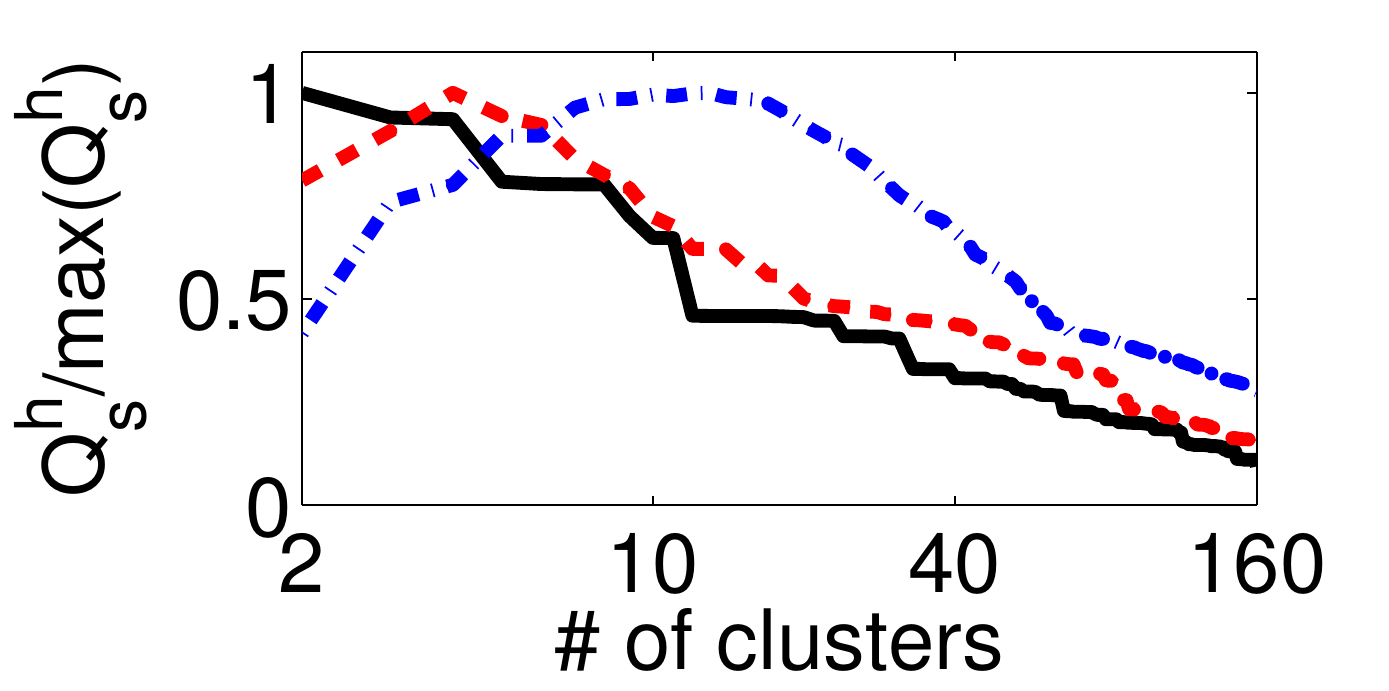}}
\end{minipage}
\hfill
\begin{minipage}[b]{0.48\linewidth}
  \centering
  \centerline{\includegraphics[width=4.0cm]{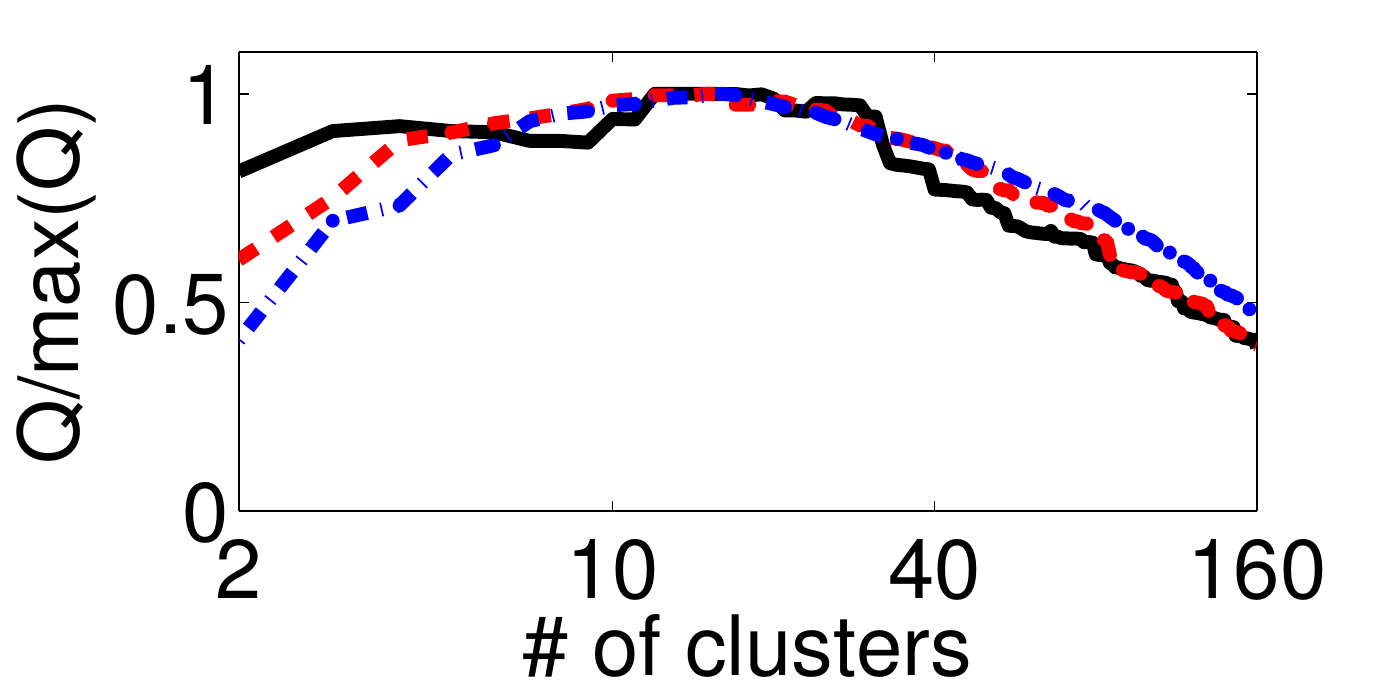}}
\end{minipage}
\caption{Modularity: scale-dependent $Q^h_s$ (left) or usual $Q$ (right) with respect to the number of clusters for three different scales $s$ ($s=1.2$ in dot-dashed blue, 
$2.4$ in dashed red, $3.3$ in black line). 
The maximum should estimate the number of communities at each scale; it is correct only for $Q^h_s$.
}
\label{fig:results_modu}
\end{figure}

A next question is: if we don't use our \emph{a priori} knowledge of the theoretical number of clusters, does the maximisation of the proposed 
scale-dependent modularity estimate the correct number of clusters?
Fig.~\ref{fig:results_modu} displays the scale-dependent (left) and classical (right) modularity with respect to the number of 
clusters kept in the dendrogram, for three different scales $s$.
The maxima that define the best partition at each scale are significantly different. For classical modularity, it is always near the same number of clusters 
(hence pointing to communities having always the same mean size). For the scale-dependent modularity, it changes with 
$s$: the smaller $s$ points to a higher number of clusters, hence smaller clusters.
Here, we used $Q^h_s$ but results are similar with wavelets.
This illustrates why we need a scale-dependent modularity.

\begin{figure}[tb]
\begin{minipage}[b]{.48\linewidth}
  \centering
  \centerline{(a) \includegraphics[width=4.0cm]{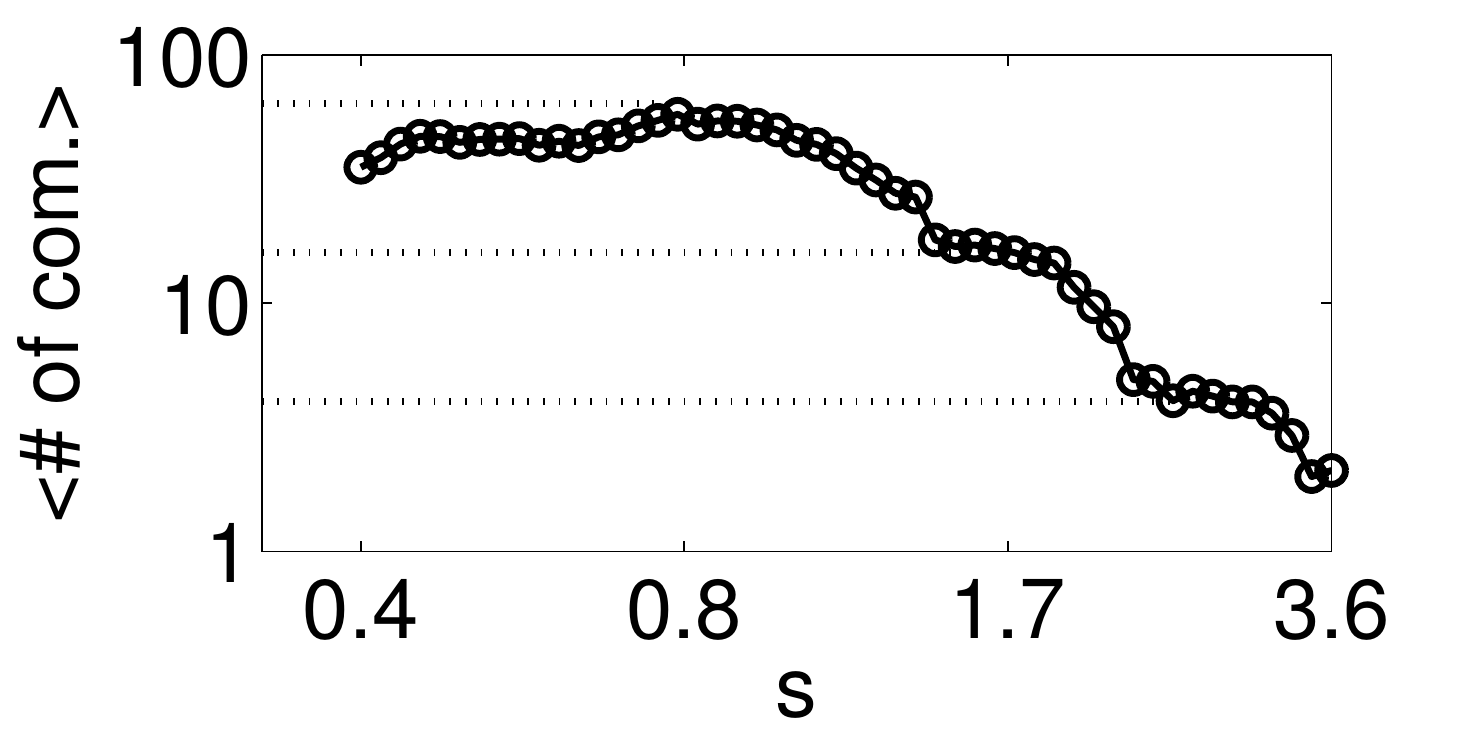}}
\end{minipage}
\hfill
\begin{minipage}[b]{0.48\linewidth}
  \centering
  \centerline{(b) \includegraphics[width=4.0cm]{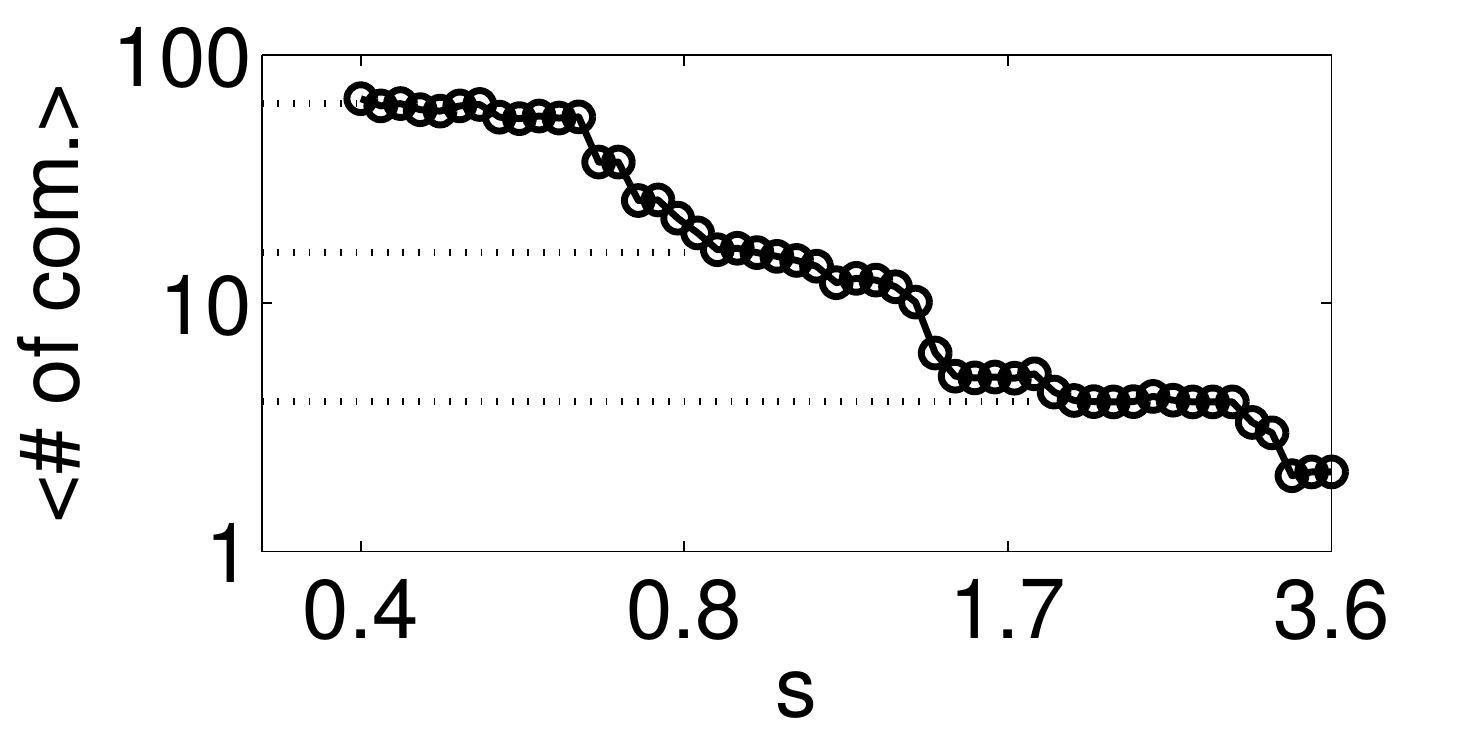}}
\end{minipage}

\begin{minipage}[b]{.48\linewidth}
  \centering
  \centerline{(c) \includegraphics[width=4.0cm]{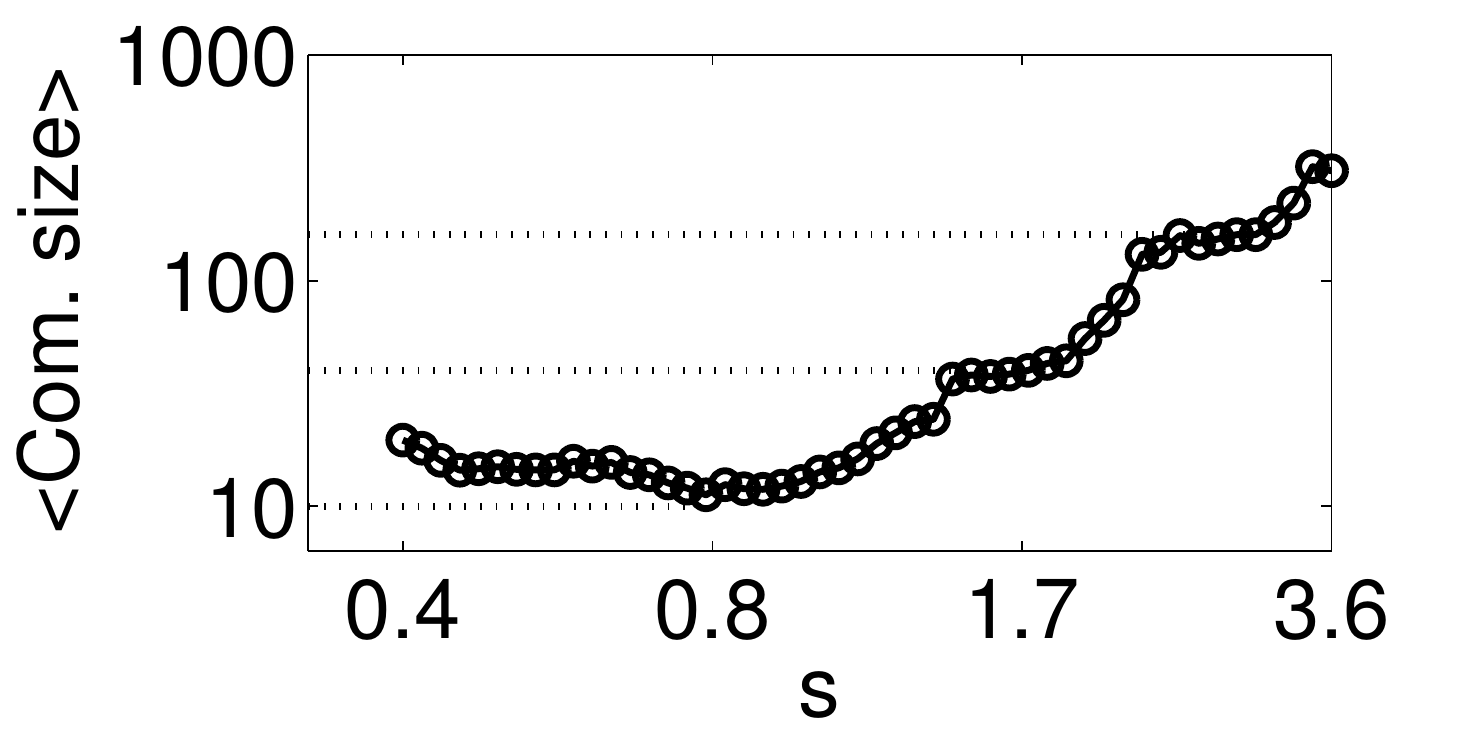}}
\end{minipage}
\hfill
\begin{minipage}[b]{0.48\linewidth}
  \centering
  \centerline{(d) \includegraphics[width=4.0cm]{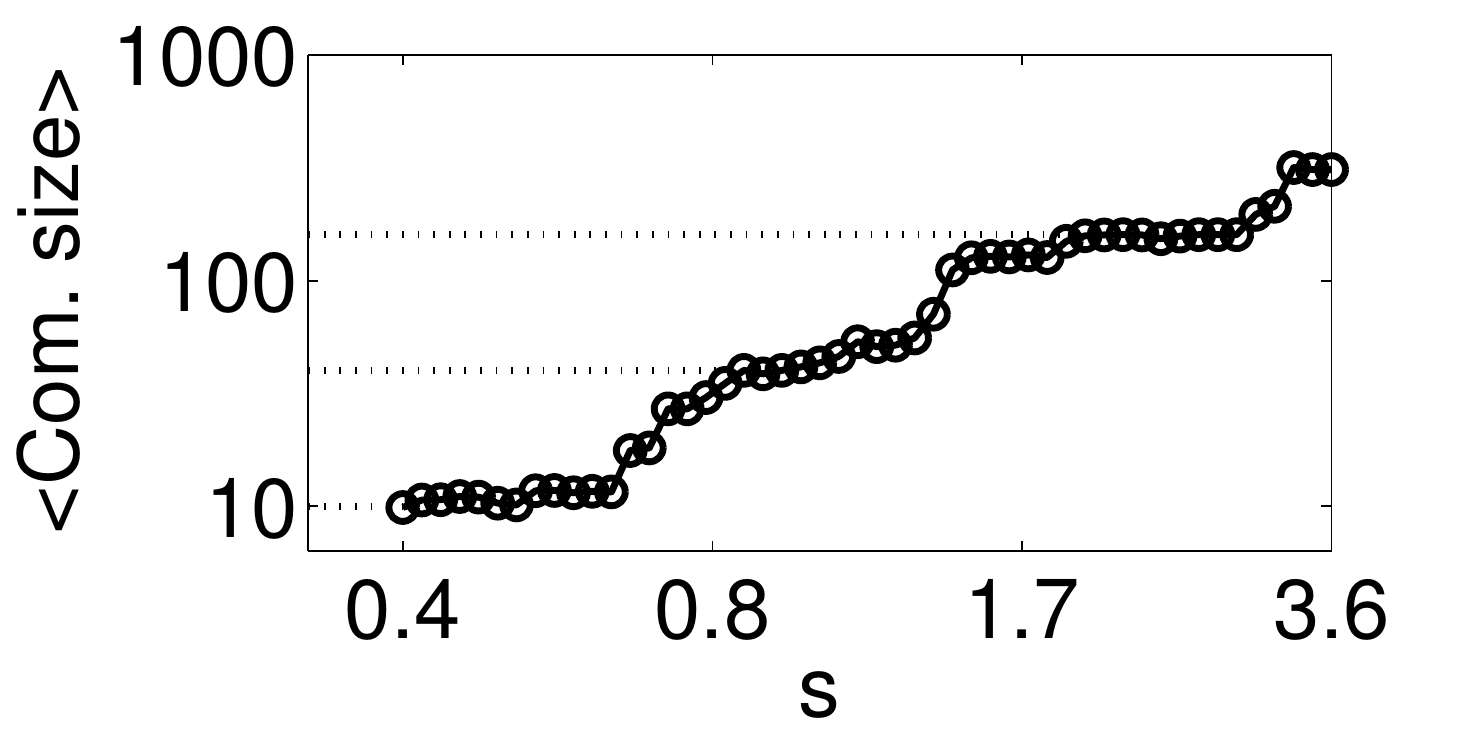}}
\end{minipage}

\caption{
(a), (b) Average (over 100 bootstrap samples of the graph) of the estimated number of communities. 
(c), (d) Average size of uncovered communities.
Left: using wavelets. Right: using scaling functions.
}
\label{fig:results2}
\end{figure}

Quantitative results for the method are obtained by creating bootstrap samples of the network model of \ref{sec:2model}, 
by randomly  adding $\pm 10\%$ of the weight of each link as in~\cite{lambiotte2010multi,rosvall2010mapping}.
We plot in Fig.~\ref{fig:results2} the average (over 100 bootstrap samples)  of the estimated number of communities (top), and the average size of the uncovered 
communities (bottom). The dotted horizontal lines correspond to the three theoretical community structures. The method successfully recovers these different community
structures, both in term of number of communities and size. 
Using scaling functions lead to better results, with larger and better defined intervals
where these numbers and sizes are correctly estimated.


\section{Conclusion}
\label{sec:concl}

We proposed a method for multiscale community mining in graphs, for which no parameter
needs to be adjusted. The mathematical soundness of graph wavelets, on which this method is based, is one of its great asset. 
The first aspect of our work was to complement certain aspects of graph wavelets
(range of scale, parameters of the band-pass filter kernel, and corresponding graph scaling functions)
for their application to community detection in graphs. 
For instance, scaling functions associated to graph wavelets provide slightly better results
for scale-dependent community mining.  

The weakness of the method is the computational cost. 
Two steps are costly: the diagonalisation of the Laplacian, and the evaluation of each possible subdivision of the dendrograms. 
Regarding the first problem, Hammond et. al~\cite{hammond2011wavelets} proposed a fast wavelet transform based 
on Chebyshev polynomial approximation that does not require the Laplacian's diagonalisation. This 
approximation will be easily used in our method.
For the dendrogram mining, instead of looking at all possible subdivisions, future work may look only with subdivisions given by the largest gaps of the dendrograms, or,
as the scale-dependent modularity has a bell shape curve (see Fig.~\ref{fig:results_modu}), one could search only for a large local maximum.



\bibliographystyle{IEEEbib}
\bibliography{biblio_EUSIPCO.bib}

\begin{thebibliography}{10}

\bibitem{Chung2006}
F.R.K. Chung and L.~Lu,
\newblock {\em Complex graphs and networks},
\newblock Number 107. Amer Mathematical Society, 2006.

\bibitem{Barrat2008}
A.~Barrat, M.~Barthlemy, and A.~Vespignani,
\newblock {\em Dynamical processes on complex networks},
\newblock Cambridge University Press, 2008.

\bibitem{Fortunato2010}
S.~Fortunato,
\newblock ``Community detection in graphs,''
\newblock {\em Physics Reports}, vol. 486, no. 3-5, pp. 75--174, 2010.

\bibitem{newman2006modularity}
M.E.J. Newman,
\newblock ``Modularity and community structure in networks,''
\newblock {\em PNAS}, vol. 103, no. 23, pp. 8577, 2006.

\bibitem{fortunato2007resolution}
S.~Fortunato and M.~Barthelemy,
\newblock ``Resolution limit in community detection,''
\newblock {\em PNAS}, vol. 104, no. 1, pp. 36, 2007.

\bibitem{kumpula2007limited}
J.M. Kumpula, J.~Saram{\"a}ki, K.~Kaski, and J.~Kertesz,
\newblock ``Limited resolution in complex network community detection with
  potts model approach,''
\newblock {\em Eur. Phys. J. B}, vol. 56, no. 1, pp. 41--45, 2007.

\bibitem{schaub2012markov}
M.T. Schaub, J.C. Delvenne, S.N. Yaliraki, and M.~Barahona,
\newblock ``Markov dynamics as a zooming lens for multiscale community
  detection: non clique-like communities and the field-of-view limit,''
\newblock {\em PloS one}, vol. 7, no. 2, pp. e32210, 2012.

\bibitem{lambiotte2010multi}
R.~Lambiotte,
\newblock ``Multi-scale modularity in complex networks,''
\newblock in {\em Proc. 8th Int. Symp. WiOpt}, 2010, pp. 546--553.

\bibitem{reichardt2006statistical}
J.~Reichardt and S.~Bornholdt,
\newblock ``Statistical mechanics of community detection,''
\newblock {\em Physical Review E}, vol. 74, no. 1, pp. 016110, 2006.

\bibitem{arenas2008analysis}
A.~Arenas, A.~Fernandez, and S.~Gomez,
\newblock ``Analysis of the structure of complex networks at different
  resolution levels,''
\newblock {\em New Journal of Physics}, vol. 10, no. 5, pp. 053039, 2008.

\bibitem{hammond2011wavelets}
D.K. Hammond, P.~Vandergheynst, and R.~Gribonval,
\newblock ``Wavelets on graphs via spectral graph theory,''
\newblock {\em Applied and Computational Harmonic Analysis}, vol. 30, no. 2,
  pp. 129--150, 2011.

\bibitem{hammond2012incorporating}
D.K. Hammond, B.~Scherrer, and A.~Malony,
\newblock ``Incorporating anatomical connectivity into eeg source estimation
  via sparse approximation with cortical graph wavelets,''
\newblock in {\em IEEE ICASSP}, 2012, pp. 573--576.

\bibitem{Narang2012}
S.K. Narang, Y.H. Chao, and A.~Ortega,
\newblock ``Graph-wavelet filterbanks for edge-aware image processing,''
\newblock in {\em IEEE SSP Workshop}, 2012, pp. 141--144.

\bibitem{sales2007extracting}
M.~Sales-Pardo, R.~Guimera, A.A. Moreira, and L.A.N. Amaral,
\newblock ``Extracting the hierarchical organization of complex systems,''
\newblock {\em PNAS}, vol. 104, no. 39, pp. 15224--15229, 2007.

\bibitem{CarmonaBook}
R.~Carmona, W.L. Hwang, and B.~Torrésani,
\newblock {\em Practical Time-Frequency analysis},
\newblock Academic Press, 1998.

\bibitem{jain1999data}
A.K. Jain, M.N. Murty, and P.J. Flynn,
\newblock ``Data clustering: a review,''
\newblock {\em ACM computing surveys (CSUR)}, vol. 31, no. 3, pp. 264--323,
  1999.

\bibitem{donetti2004detecting}
L.~Donetti and M.A. Munoz,
\newblock ``Detecting network communities: a new systematic and efficient
  algorithm,''
\newblock {\em J. Stat. Mech.}, vol. 2004, pp. P10012, 2004.

\bibitem{danon2005comparing}
L.~Danon, A.~Diaz-Guilera, J.~Duch, and A.~Arenas,
\newblock ``Comparing community structure identification,''
\newblock {\em J. Stat. Mech.}, vol. 2005, pp. P09008, 2005.

\bibitem{rosvall2010mapping}
M.~Rosvall and C.T. Bergstrom,
\newblock ``Mapping change in large networks,''
\newblock {\em PloS one}, vol. 5, no. 1, pp. e8694, 2010.

\end{thebibliography}

\end{document}